\documentclass{ifacconf}

\usepackage{graphicx} 
\usepackage{amsmath,amssymb}
\usepackage{threeparttable,tablefootnote}
\usepackage{booktabs,makecell}
\usepackage{natbib}   

\usepackage[algo2e,noline]{algorithm2e}
\SetInd{0.8em}{0.1em}
\makeatletter
\renewcommand{\@algocf@post@ruled}{%
  \hrule height \algoheightrule\relax 
}
\makeatother

\SetAlgoInsideSkip{medskip} 
\RestyleAlgo{ruled} 

\SetKwComment{Comment}{/* }{ */}

\begin{document}
\begin{frontmatter}

\makeatletter
\vbox to 0pt{\vspace{-25pt}\noindent\hspace*{0pt}{\footnotesize Accepted for publication in the 23rd IFAC World Congress (IFAC WC 2026), Busan, Republic of Korea.}\vss}
\makeatother
\vspace*{-10pt}

\title{Recursive Identification of EIV-ARX Models for Time Varying SISO Processes}


\author[First]{Deepanjhan Das} 
\author[Second]{Shankar Narasimhan} 

\address[First]{Department of Chemical Engineering, Indian Institute of Technology Madras, (e-mail: deepanjhan.iitm22@gmail.com)}
\address[Second]{Department of Chemical Engineering, Indian Institute of Technology Madras, (e-mail: naras@iitm.ac.in)}

\begin{abstract} 
This paper proposes a recursive algorithm, rARX-DIPCA, for identifying errors-in-variables autoregressive models with exogenous input (EIV-ARX), for tracking time-varying SISO processes. Building on a recently developed recursive iterative PCA method, the proposed algorithm recursively updates model parameters and noise variances as new measurements arrive, without storing historical data beyond a specified lag window. The method enables real-time adaptation to sensor degradation, and changes in model coefficients. The algorithm simultaneously identifies process order, time delay, and noise variances while maintaining computational efficiency through online covariance updates. Simulation studies on benchmark systems demonstrate effective tracking performance and practical applicability.
\end{abstract}


\end{frontmatter}

\section{Introduction}
Linear dynamic model identification is fundamental to modern process system engineering, with applications spanning from soft sensing \citep{Cao:2018}, process optimization, model predictive control and fault diagnosis \citep{Prakash:2005}. Among various model structures, the autoregressive with exogenous input (ARX) model for single input and single output (SISO) systems, is preferred due to its mathematical tractability and ability to approximate complex dynamics. 

\begin{figure}[!tb]
    \centering
    \includegraphics[width=8.4cm]{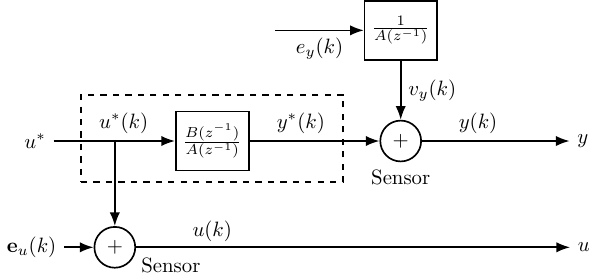}
    \caption{Linear dynamic EIV-ARX model architecture}
    \label{Figure_1}
\end{figure}

Unlike the classical framework, errors-in-variables (EIV) modeling assumes that both exogenous input and output measurements are corrupted by noise, as illustrated in Fig.~\ref{Figure_1}.  The EIV modeling framework is particularly relevant when the input and output data are directly obtained from some operating processes rather than through designed experiments. From a theoretical standpoint, this framework shares deep structural connections with dynamic generalized factor analysis, where latent true variables must be similarly estimated from noisy observation streams \citep{Picci:2023}.

Several approaches have been proposed for identifying EIV-ARX models for SISO systems.  These include methods based on bias compensation \citep{Ikenoue:2005}, bias elimination \citep{Zheng:2002}, and the dynamic Frisch scheme \citep{Diversi:2010}. All of these methods obtain consistent estimates of the ARX model parameters along with the input and output noise variances.  However, they assume the process order to be known or specified. The Instrument Variable (IV) approach \citep{Soderstrom:2002} consistently estimates the model parameters without the need to estimate the noise variances.  Recently \cite{Maurya:2022} proposed a modified dynamic iterative PCA approach that jointly estimates the process order, time delay, noise variances, and model parameters without requiring prior knowledge from the user. This technique builds on the basic Koopmans-Levin approach \citep{Fernando:1985}, by including an iterative procedure to estimate noise variances, and a theoretical criterion for estimating the order of the process based on analysis of the eigenvalues of the covariance matrix.

Most industrial processes exhibit time-varying characteristics due to catalyst deactivation, equipment fouling, or changing operating conditions. For such systems, \cite{Zhao:2018} addressed the specific challenge of robust identification of output error (OE) ARX models with time-varying time-delays using a Variational Bayesian (VB) approach, whereas \cite{Zhang:2022} developed an expectation-maximization (EM) framework to identify EIV-ARX models with time-varying time-delays. Their works demonstrate the importance of adapting to time-varying system parameters, though their approaches remain in batch processing domain and assume an a priori knowledge of process order. Batch methods become computationally prohibitive for long data sequences, and cannot adapt to more general time-varying processes.

The need for adaptive identification has motivated the development of recursive approaches, such as continuous adaptation of ordinary least squares (OLS) regression models, PCA based model \citep{Li:2000}, reduced-order models \citep{Prakash:2025}. However, these methods make simplistic assumptions about the noise variances that eliminates the requirement of knowing changes in noise variances to perform model updates. Most recently, \cite{Pradeep:2025} introduced a recursive iterative PCA (RIPCA) framework for updating static (steady-state) EIV models and unknown, heteroskedastic noise variances online without storing past data. 

In this paper, we extend the RIPCA method by drawing ideas from the modified dynamic iterative PCA approach \citep{Maurya:2022} appropriately to handle dynamic structure and colored noise characteristics of EIV-ARX models. We thereby provide an efficient recursive identification approach that is capable of continually adapting to changes in heteroskedastic noise variances, and coefficients of EIV-ARX models.

The remainder of this paper is structured as follows. In Section~\ref{sec:foundation}, we provide a formal description of the EIV identification problem of SISO ARX processes and the modified dynamic iterative PCA method. This forms the basis of the proposed recursive identification procedure of EIV-ARX models for time varying processes, which we present in Section~\ref{sec:method}. In Section~\ref{sec:app-study}, we demonstrate the effectiveness of the proposed algorithm in adapting to slow sensor degradation and changing process parameters using two simulation studies. Section~\ref{sec:conclusion} addresses the concluding remarks and potential future extensions.

\section{Foundations}\label{sec:foundation}

\subsection{Dynamic Iterative PCA for identifying EIV-ARX models of SISO processes} \label{subsec:DIPCA}
In this section, we briefly discuss the dynamic iterative PCA (DIPCA) method for identifying an EIV-ARX model, since this forms the basis for the proposed recursive EIV-ARX model identification approach.  

The deterministic linear time-invariant (LTI) SISO discrete dynamical model between noise-free input ($u^*$) and output ($y^*$) variables can in general be described as
\begin{equation}
    \label{eq:2.3}
    y^*(k) = \sum_{i=1}^{n_y} a_iy^*(k-i) + \sum_{j=D}^{n_u} b_ju^*(k-j)
\end{equation}
where $D, n_u, n_y$ are input-output delay, input and output orders, respectively. The process order is defined as $\eta = \max(n_u, n_y)$. In the case of an EIV-ARX process, as shown in Fig.~\ref{Figure_1}, where $A(z^{-1}) = 1 - \sum_{i=1}^{n_y} a_iz^{-1}$ and $B(z^{-1}) = \sum_{j=D}^{n_u} b_jz^{-1}$, both the input and output measurements contain additive noise which are modeled as follows:
\begin{subequations}
    \label{eq:2.4}
    \begin{align}
        &y(k) = y^*(k) + v_y(k); \quad e_y(k) \sim \mathcal{N}\left(0, \sigma_{e_y}^2\right) \\
        &v_y(k) = \sum_{i=1}^{n_y} a_iv_y(k-i) + e_y(k);  \\
        &u(k) = u^*(k) + e_u(k); \quad e_u(k) \sim \mathcal{N}\left(0, \sigma_{e_u}^2\right)
    \end{align}
\end{subequations}
where $e_y$ and $e_u$ are assumed to be mutually independent. The comprehensive identification problem requires estimates of parameters $D, n_u, n_y, \sigma_{e_u},$ $\sigma_{e_y},$ $\{a_i\}_{i=1}^{n_y},$ and $\{b_j\}_{j=D}^{n_u}$ to be obtained from $N$ noisy measured samples.

Equation (\ref{eq:2.3}) can be recast as 
\begin{equation}
    \label{eq:2.1}
\boldsymbol{\theta}^\intercal\mathbf{z}_{\eta}^*(k) = 0
\end{equation}
where $\boldsymbol{\theta}$ contains the model parameters and $\mathbf{z}_\eta^*(k)$ contains $\eta$ lagged noise-free variables defined as follows. 
\begin{subequations}
    \label{eq:2.5a}
    \begin{align}
        &\boldsymbol{\theta}(k) = \begin{bmatrix}
                1 & a_1 & \ldots & a_{\eta} & b_1 & \ldots & b_{\eta}  
            \end{bmatrix}^{\intercal} \\
        &\mathbf{z}^*_{\eta}(k) = \begin{bmatrix}
                y^*(k) & \ldots & y^*(k-\eta) & u^*(k) & \ldots & u^*(k-\eta)     
            \end{bmatrix}^{\intercal}
    \end{align}
\end{subequations}
Since $\eta$ and the true values are unknown, a lagged data matrix $\mathbf{Z}_L$ is constructed using a sufficiently large lag window $L \geq \eta$, to identify the model parameters.
\begin{subequations}
    \label{eq:2.5}
    \begin{align}
        &\mathbf{z}_L(k) = \begin{bmatrix}
                y(k) & \ldots & y(k-L) & u(k) & \ldots & u(k-L)     
            \end{bmatrix}^\intercal \\
        &\mathbf{Z}_L = \begin{bmatrix}
                \mathbf{z}_L(L+1) & \mathbf{z}_L(L+2) & \ldots & \mathbf{z}_L(N)
            \end{bmatrix}^\intercal
    \end{align}
\end{subequations}
Equations (\ref{eq:2.3}) to (\ref{eq:2.5}) can be written in compact form as
\begin{subequations}
    \label{eq:2.6}
    \begin{align}
        &\mathbf{A}_d\left(\mathbf{Z}_L^*\right)^\intercal = \begin{bmatrix} \mathbf{A}_y & \mathbf{B}_u \end{bmatrix} \left(\mathbf{Z}_L^*\right)^\intercal = \mathbf{0}_{d\times (N-L)} \\
        &\mathrm{subject\ to}:\quad \mathbf{Z}_L = \mathbf{Z}_L^* + \mathbf{E}_L
    \end{align}
\end{subequations}
where each row of the constraint matrix $\mathbf{A}_d: d\times 2(L+1)$ is a time-shifted version of the model coefficient vector $\boldsymbol{\theta}$. The use of a lag window greater than $\eta$ results in $d = L-\eta+1$ linearly independent equations relating the true lagged variables $\mathbf{z}^*_L$. 

For the EIV-ARX model, the covariance matrix of the noise corrupting the lagged data vector $\mathbf{z}_L(k)$ denoted as $\mathbf{\Sigma}_{\mathbf{e}_L}$ is given by
\begin{equation}
    \label{eq:2.7}
    \mathbf{\Sigma}_{\mathbf{e}_L} = \begin{bmatrix}
            \mathbf{\Sigma}_{\mathbf{v}_y} & \mathbf{0} \\
            \mathbf{0} & \sigma^2_{e_u}\mathbf{I}_{L+1}
        \end{bmatrix} 
\end{equation}
where $\mathbf{\Sigma}_{\mathbf{v}_y}$ is a Toeplitz matrix with elements $\gamma_v(\ell)$, which is the ACVF of $v_y(\ell)$ at lag $\ell$. Given $\sigma_{e_y}$ and $\{a_i\}_{i=1}^L$, the ACVF is estimated by solving the Yule-Walker equations
\begin{equation}
    \label{eq:2.8}
    \gamma_v(\ell) - \sum_{i=1}^L a_i\gamma_v(\ell-i) = \begin{cases}
        \sigma^2_{e_y}, & \ell = 0 \\
        0, & \ell > 0
    \end{cases}
\end{equation}

\cite{Maurya:2022} developed a method for identifying an EIV-ARX model of a SISO process based on modification of the dynamic iterative PCA approach \citep{Maurya:2018}, which is capable of automatically estimating the noise variances, order and delay of the process, and ARX model coefficients. The method applies PCA to the lagged data matrix after scaling it using the Cholesky factor of corresponding noise covariance matrix $\mathbf{\Sigma}_{\mathbf{e}_L}$. The scaled lagged data matrix is therefore given by 
\begin{equation}
    \label{eq:2.9}
    {\mathbf{Z}_{L,s}} = {\mathbf{Z}_{L}}\mathbf{L}_{\mathbf{e}_L}^{-1} 
\end{equation}
where
\begin{equation}
    \label{eq:2.9b}
    \mathbf{\Sigma}_{\mathbf{e}_L} = \mathbf{L}_{\mathbf{e}_L}\mathbf{L}_{\mathbf{e}_L}^{\intercal}
\end{equation}

It can be shown that the smallest $d$ singular values of the scaled data matrix ${\mathbf{Z}_{L,s}}$ should all be equal to unity, in the limit as $N$ tends to infinity.  This property is exploited to estimate the number of constraints (and consequently estimate the process order) by applying a hypothesis test to assess the equality of the smallest singular values. The constraint matrix $\mathbf{A}_d$ is estimated from the right singular vectors $\mathbf{V}_d$ corresponding to the $d$ unity singular values as
\begin{equation}
    \label{eq:2.9a}
    \mathbf{\hat{A}}_d = \mathbf{V}_d^\intercal \mathbf{L}_{\mathbf{e}_L}^{-1}
\end{equation}

In order to estimate the unknown noise variances, the above model estimation is combined with a method for estimating the noise variances within an iterative scheme. The two noise variances $\sigma^2_{e_y}$ and $\sigma^2_{e_u}$ are estimated by minimizing the following joint likelihood function of the lagged constraint residuals
\begin{equation}
    \label{eq:2.10}
    \begin{array}{ll}
        \underset{\sigma^2_{e_y},\ \sigma^2_{e_u}}{\mathrm{min}}\ \Biggl[ (N-L) &\mathrm{log} \left| \mathbf{\Sigma}_{\mathbf{r}_L} \right|\ + \\ &\sum_{k=L+1}^N \mathbf{r}_L(k)^\intercal \left( \mathbf{\Sigma}_{\mathbf{r}_L} \right)^{-1} \mathbf{r}_L(k) \Biggr]
    \end{array}
\end{equation}
where, $\mathbf{r}_L(k) = \mathbf{\hat{A}}_d\mathbf{z}_L(k)$ and  $\mathbf{\Sigma}_{\mathbf{r}_L} = \mathbf{\hat{A}}_d \mathbf{\hat{\Sigma}}_{\mathbf{e}_L} \mathbf{\hat{A}}_d^\intercal$.  It may be noted from (\ref{eq:2.7}) and (\ref{eq:2.8}), that estimates of the ARX model coefficients are required for computing $\mathbf{\Sigma}_{\mathbf{e}_L}$. Due to rotational ambiguity, $\mathbf{\hat{A}}_d,$ estimated using (\ref{eq:2.9a}) is related to $\mathbf{A}_d$ via a non-singular rotation matrix $\mathbf{R}$ as $\mathbf{A}_d = \mathbf{R\hat{A}}_d$. \cite{Maurya:2022} have provided a formal procedure to uniquely determine $\mathbf{R}$ for obtaining the estimates of the ARX model coefficients, by exploiting the fact that the zeros and unity elements of the matrix $\mathbf{A}_d$ are known. The overall algorithm consists of three nested iterative loops.  An outer loop where the number of constraints $d$ is guessed and decremented in each iteration until the smallest $d$ singular values corresponding to the scaled data matrix ${\mathbf{Z}_{L,s}}$ are equal, an inner loop in which the model and noise variances are alternatively estimated, and an innermost loop to solve the minimization problem defined by (\ref{eq:2.10}). The reader is referred to \cite{Maurya:2022} for a detailed description of the overall modified dynamic iterative PCA algorithm for identifying an EIV-ARX model from a given sample of measurements. 

\section{Proposed Methodology}\label{sec:method}
In this section, we describe the proposed recursive estimation method to identify time-varying EIV-ARX processes by extending the modified DIPCA method described in the preceding section. Recently a recursive iterative PCA method (RIPCA) was developed to identify time varying constraint models for steady state processes \citep{Pradeep:2025}. In a recursive setting, the full data matrix is not stored or available. However, the first and second order statistics of the data can be recursively updated online.  Furthermore, the eigenvectors of the data covariance matrix span the same subspace as the right singular vectors of the data matrix, which are required to estimate the constraint matrix.  We propose to use a similar strategy to develop the recursive version of EIV-ARX model identification. Two modifications are made to the DIPCA algorithm described in the preceding section so that it only uses the covariance matrix of lagged measurements, rather than the lagged data matrix.  

In the constraint model $\mathbf{A}_d$ estimation step, the eigenvectors of the sample covariance matrix of the scaled lagged data are computed using eigen-decomposition, instead of using the singular value decomposition (SVD) of the scaled lagged data matrix. The sample covariance matrix of scaled lagged data at any time instant $k$ is given by
\begin{equation}
    \label{eq:3.2}
    \mathbf{S}_{\mathbf{Z}_{L,s},k} =\frac{1}{N_k} \mathbf{L}_{\mathbf{e}_L}^{-\intercal}{\mathbf{Z}^{\intercal}_{L,k}}{\mathbf{Z}_{L,k}}\mathbf{L}_{\mathbf{e}_L}^{-1}
\end{equation}
where, $N_k$ is the number of samples till the $k$'th instant.

The second modification is to recast the optimization problem defined in (\ref{eq:2.10}) to update the noise variances $\hat{\sigma}_{e_y,k-1}^2,$ and $\hat{\sigma}^2_{e_u,k-1}$, so that it also requires only the sample covariance matrix of lagged  measurements rather than the data matrix. By applying the cyclic property of the trace operator in (\ref{eq:2.10}) \citep{Pradeep:2025}, the optimization problem can be rewritten as
\begin{equation}
    \label{eq:3.3}
    \begin{array}{ll}
        \mathbf{\hat{\Sigma}}_{\mathbf{e}_L,k} &= \mathrm{arg}\ \underset{\sigma^2_{e_y},\ \sigma^2_{e_u}}{\mathrm{min}} \Biggl[ (N-L)\ \mathrm{log}\left| \mathbf{\Sigma}_{\mathbf{r}_{L,k}} \right|\ + \\ &(N-L)\ \mathrm{Tr} \left( \left( \mathbf{\Sigma}_{\mathbf{r}_{L,k}} \right)^{-1} \mathbf{\hat{A}}_{d,k} \mathbf{{S}}_{\mathbf{Z}_{L},k} \mathbf{\hat{A}}_{d,k}^\intercal \right) \Biggr] 
    \end{array}
\end{equation}
where, $\mathbf{\Sigma}_{\mathbf{r}_{L,k}} = \mathbf{\hat{A}}_{d,k} \mathbf{{\hat \Sigma}}_{\mathbf{e}_{L},k-1} \mathbf{\hat{A}}_{d,k}^\intercal$. It may be noted that the above optimization problem only consists of the noise variances of input and output measurements as decision variables, since the noise covariance matrix of lagged data $\mathbf{\hat{\Sigma}}_{\mathbf{e}_L,k}$ can be computed from these noise variances using (\ref{eq:2.7}) and (\ref{eq:2.8}). Similar to DIPCA, the above two steps are iteratively solved to estimate the constraint model $\mathbf{\hat{A}}_{d,k},$ and noise variances $\hat{\sigma}_{e_y,k}^2,$ $\hat{\sigma}^2_{e_u,k}$.

The sample covariance matrix of the lagged data $\mathbf{{S}}_{\mathbf{Z}_{L},k}$ corresponding to the $k$'th instant can be computed in a recursive manner when new measurements are received at any time instant $k$. This only requires (a) the measurements corresponding to the preceding $L$ time instances, and (b) the sample covariance matrix of lagged measurements at the preceding time instant, i.e., $\mathbf{{S}}_{\mathbf{Z}_{L},k-1}$ to be stored. The recursive equation for computing the sample covariance matrix of lagged data is given by
\begin{equation}
    \label{eq:3.1}
    \mathbf{{S}}_{\mathbf{Z}_{L},k} = \frac{N_{k-1}}{N_k} \mathbf{{S}}_{\mathbf{Z}_{L},k-1} + \frac{1}{N_k}\mathbf{z}_{L}(k) \mathbf{z}^\intercal_{L}(k)
\end{equation}
where $N_{k-1}$ is the number of samples obtained before the $k$'th instant, and $N_k = N_{k-1} + 1$.

In the above description, the proposed method is applied at every time instant to re-estimate all the model parameters.  In practice, while the covariance matrix can be updated at every time instant, the model re-identification can be triggered on need basis based on monitoring the model prediction errors.  This can significantly reduce the computational burden. Secondly, the covariance matrix update using (\ref{eq:3.1}), as a part of the baseline recursive algorithm presented here, assigns equal weight to all measurements. A forgetting factor can optionally be introduced as a hyperparameter by practitioners to give more weight to recent measurements as described by \cite{Li:2000}, which would result in faster convergence to the new system parameters. The proposed method, referred to as rARX-DIPCA is summarized in Algorithm \ref{algo:1}.

\SetKwFor{While}{while}{do}{end{\hspace{0.25em}}while}
\SetKwInput{KwRequire}{Prior Information}
\SetKwInput{KwParam}{Configuration Parameters}
\newcommand{\hangKwText}[1]{\hangindent=3.25em\hangafter=1 #1}
\begin{algorithm2e}[!hbtp]
    \begin{minipage}{\dimexpr8.34cm-0\algomargin\relax}
    \caption{Recursive adaptation of time-varying EIV-ARX processes using the rARX-DIPCA algorithm}
    \label{algo:1}
    \vspace{-0.5em}
    \KwIn{\hangKwText{The data $\mathbf{z}_{L}(k),$ and lag $L$}}
    \KwRequire{\hangKwText{From $(k-1)$'th instant, $\mathbf{{S}}_{\mathbf{Z}_{L},k-1}$, and $N_{k-1}$, as well as the estimates of $\mathbf{\hat{a}}_{L} =$ $\{\hat{a}_j\}_{j=1}^{L}$, $\hat{\sigma}_{e_y,k-1},$ $\hat{\sigma}_{e_u,k-1}$}}
    \KwParam{\hangKwText{Max iterations $i_{max} > 0,$ stopping tolerance $\varepsilon_\lambda > 0$}}
    \KwOut{\hangKwText{Updated estimates of $\mathbf{\hat{a}}_{\hat{\eta}_k} = \{\hat{a}_j\}_{j=1}^{\hat{\eta}_k},$ $\mathbf{\hat{b}}_{\hat{\eta}_k} = \{\hat{b}_j\}_{j=0}^{\hat{\eta}_k},$ $\hat{\sigma}_{e_y,k},$ and $\hat{\sigma}_{e_u,k}$}}
    \vspace{0.5em}
        \nl\label{algo1:st1}Compute $N_k = N_{k-1} + 1$. Initialize $i \leftarrow 1,$ $\lambda^{(0)} \leftarrow 0,$ $\lambda^{(i)} \leftarrow 1,$ $\hat{\sigma}_{e_y,k}^{(i)} \leftarrow \hat{\sigma}_{e_y,k-1},$ $\hat{\sigma}_{e_u,k}^{(i)} \leftarrow \hat{\sigma}_{e_u,k-1}$\;

        \nl\label{algo1:st2}Obtain initial estimate of $\mathbf{\hat{\Sigma}}_{\mathbf{e}_L,k}^{(i)}$ using $\mathbf{\hat{a}}_L,$ $\hat{\sigma}_{e_y,k}^{(i)},$ and $\hat{\sigma}_{e_u,k}^{(i)}$ based on (\ref{eq:2.7}) and (\ref{eq:2.8})\;

        \nl\label{algo:st3}Compute $\mathbf{{S}}_{\mathbf{Z}_{L},k}$ using $\mathbf{{S}}_{\mathbf{z}_{L},k-1}$ and $\mathbf{z}_{L}(k)$ as in (\ref{eq:3.1})\;

        \nl\label{algo1:st4}\While{$i \leq i_{max}$ $\mathbf{and}$ $\left|\left(\lambda^{(i)} - \lambda^{(i-1)}\right) / \lambda^{(i-1)}\right| > \varepsilon_\lambda$} {
            \nl\label{algo1:st5}Scale the covariance using (\ref{eq:3.2}) to obtain $\mathbf{S}_{\mathbf{Z}_{L,s},k}^{(i)}$ using $\mathbf{\hat L}_{\mathbf{e}_{L,k}}^{(i)}$, the Cholesky factor of $\mathbf{\hat{\Sigma}}_{\mathbf{e}_L,k}^{(i)}$\;

            \nl\label{algo1:st6}$[\mathbf{\hat V}\ \mathbf{\hat D}] = \mathrm{eig}\left( \mathbf{S}_{\mathbf{Z}_{L,s},k}^{(i)} \right)$. Obtain $\hat{d}$ by employing hypothesis test on $\mathbf{\hat D}$ as detailed in \cite{Pradeep:2025}.  The estimated process order is $\hat{\eta}_k = L - \hat{d}+1$\;

            \nl\label{algo1:st7}Compute $\mathbf{\hat{A}}_{\hat d,k}^{(i+1)} = \mathbf{\hat V}_{\hat d}^\intercal \times \mathbf{\hat L}_{\mathbf{e}_{L,k}}^{(i)-1}$ as in (\ref{eq:2.9a})\;

            \nl\label{algo1:st8}Obtain the rotation matrix $\mathbf{R}$ by leveraging the known structure of $\mathbf{A}_d$. $\mathbf{\hat{A}}_{\hat d,k}^{(i+1)} \leftarrow \mathbf{R}\times\mathbf{\hat{A}}_{\hat d,k}^{(i+1)}$\;

            \nl\label{algo1:st9}Obtain the updated estimates of $\hat{\sigma}^{(i+1)}_{e_y,k}, \hat{\sigma}^{(i+1)}_{e_u,k}$ by solving (\ref{eq:3.3}) using $\mathbf{{S}}_{\mathbf{Z}_{L},k},$ $\mathbf{\hat{A}}_{\hat d,k}^{(i+1)},$ $\hat{\sigma}^{(i)}_{e_y,k},$ and $\hat{\sigma}^{(i)}_{e_u,k}$\;

            \nl\label{algo1:st10}Compute $\mathbf{\hat{\Sigma}}_{\mathbf{e}_L,k}^{(i+1)}$ using $\mathbf{\hat{A}}_{\hat d,k}^{(i+1)},$ $\hat{\sigma}^{(i+1)}_{e_y,k},$ and $\hat{\sigma}^{(i+1)}_{e_u,k}$ based on (\ref{eq:2.7}) and (\ref{eq:2.8})\;

            \nl\label{algo1:st11}$\lambda^{(i-1)} \leftarrow \lambda^{(i)},$ $i \leftarrow i+1$\; 
            
            \nl\label{algo1:st12}Compute $\lambda^{(i)}$ as sum of the $\hat{d}$ smallest eigenvalues contained in $\mathbf{\hat{D}}$\;
        }
        \nl\label{algo1:st13}Set $\hat{\sigma}_{e_y,k} \leftarrow \hat{\sigma}^{(i+1)}_{e_y,k},$ $\hat{\sigma}_{e_u,k} \leftarrow \hat{\sigma}^{(i+1)}_{e_u,k}$. Extract the coefficients $\mathbf{\hat{a}}_{\hat{\eta}_k},$ and $\mathbf{\hat{b}}_{\hat{\eta}_k}$ from $\mathbf{\hat{A}}_{\hat d,k}^{(i+1)}$\;
    \end{minipage}
\end{algorithm2e}

\section{Applications}\label{sec:app-study}
We now demonstrate the performance of rARX-DIPCA through two simulation studies based on two systems, drawn from \cite{Maurya:2022}, to illustrate its application to modeling time-varying systems in two specific engineering scenarios. In each of the simulation studies, we compute the following two metrics after each sample is received for evaluating the algorithm's performance
\begin{enumerate}
\item The absolute difference between the estimated and true process orders, computed as $|\hat{\eta} - \eta|$.
\item The sum of the relative difference between the estimated and true noise standard deviations for both input and output variables, computed as   
\begin{equation}
    \label{eq:4.0}
    \left| \frac{\hat{\sigma}_{e_y,k} - \sigma_{e_y}}{\sigma_{e_y}} \right| + \left| \frac{\hat{\sigma}_{e_u,k} - \sigma_{e_u}}{\sigma_{e_u}} \right|
\end{equation}
\end{enumerate}
In addition, we also compare the estimated model coefficients obtained at the final time instant with their corresponding true values.

\subsection{Tracking Sensor Degradation} \label{subsec:study1}
\begin{figure}[!htbp]
    \begin{center}
    \includegraphics[width=8.4cm]{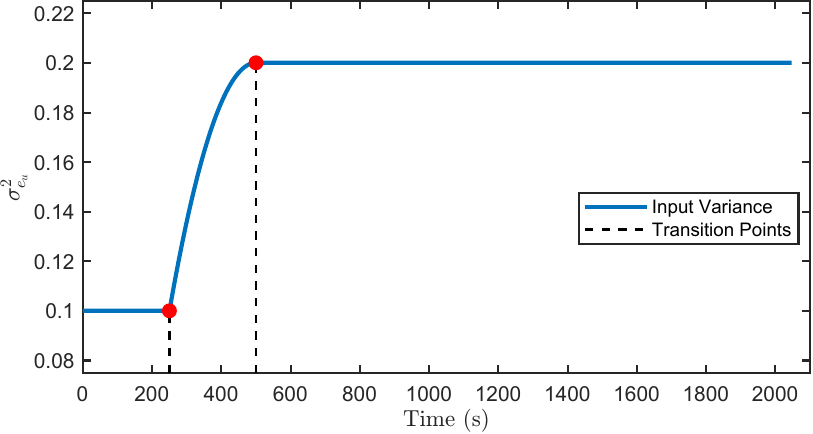}    
    \caption{Slow degradation in the input noise variance from $250$'th instant up to $500$'th instant.} 
    \label{Figure_2}
    \end{center}
\end{figure}

\begin{figure}[!htbp]
    \begin{center}
    \includegraphics[width=8.4cm]{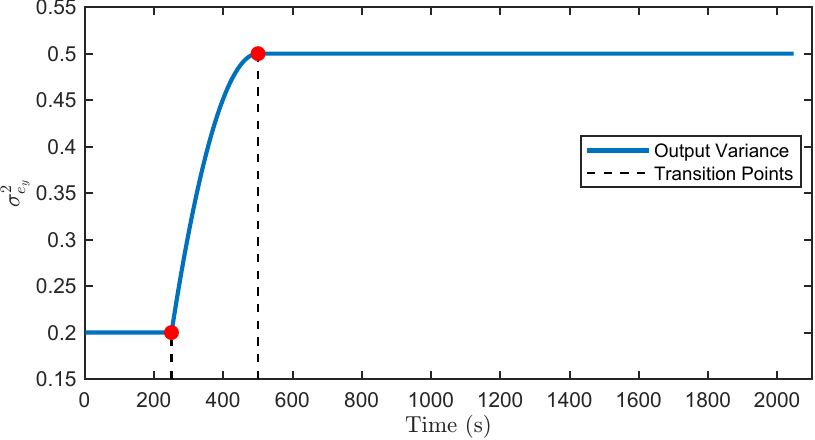}    
    \caption{Slow degradation in the output noise variance from $250$'th instant up to $500$'th instant.} 
    \label{Figure_3}
    \end{center}
\end{figure}

We consider the following second-order LTI system
\begin{equation}
    \label{eq:4.1}
    \begin{array}{ll}
        y^*(k) =& 1.5y^*(k-1) - 0.7y^*(k-2)\ + \\ & u^*(k-1) + 0.5u^*(k-2)
    \end{array}
\end{equation}
where, we choose full band random binary signal (RBS) as input $u^*$ with sample size $2047$, and generate the noise-free output $y^*$ using (\ref{eq:4.1}). We generate the noisy measurements by adding noise to the true values following the ARX structure, defined in (\ref{eq:2.4}). To simulate degradation in sensor precision, the noise variances $\sigma^2_{e_u}, \sigma^2_{e_y}$ are kept constant for the first $250$ samples equal to $0.1$ and $0.2$, respectively, maintaining a signal-to-noise ratio (SNR) of $10$. The noise variances are subsequently gradually increased in a quadratic fashion up to the $500$'th instant until $\sigma^2_{e_u} = 0.2,$ and $\sigma^2_{e_y} =0.5$ after which they are kept constant for the remaining samples. The variation in the simulated noise variances are reported in Figs.~\ref{Figure_2} and \ref{Figure_3}.

The initial estimates required to kick start rARX-DIPCA are obtained by applying the modified dynamic iterative PCA approach discussed in Section~\ref{subsec:DIPCA}, on the first $100$ samples using a lag $L=8$. At each subsequent time instant when we receive a new measurement sample, rARX-DIPCA is used to re-estimate the noise variances, process order, and model coefficients.

\begin{figure}[!htbp]
    \begin{center}
    \includegraphics[width=8.4cm]{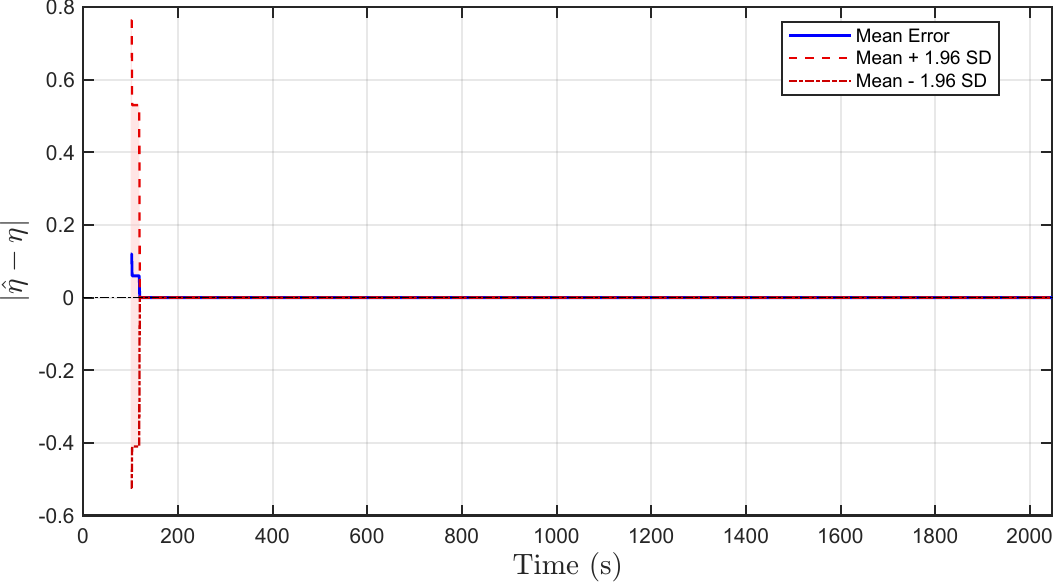}    
    \caption{Absolute difference between the estimated and true process orders.} 
    \label{Figure_4}
    \end{center}
\end{figure}

\begin{figure}[!htbp]
    \begin{center}
    \includegraphics[width=8.4cm]{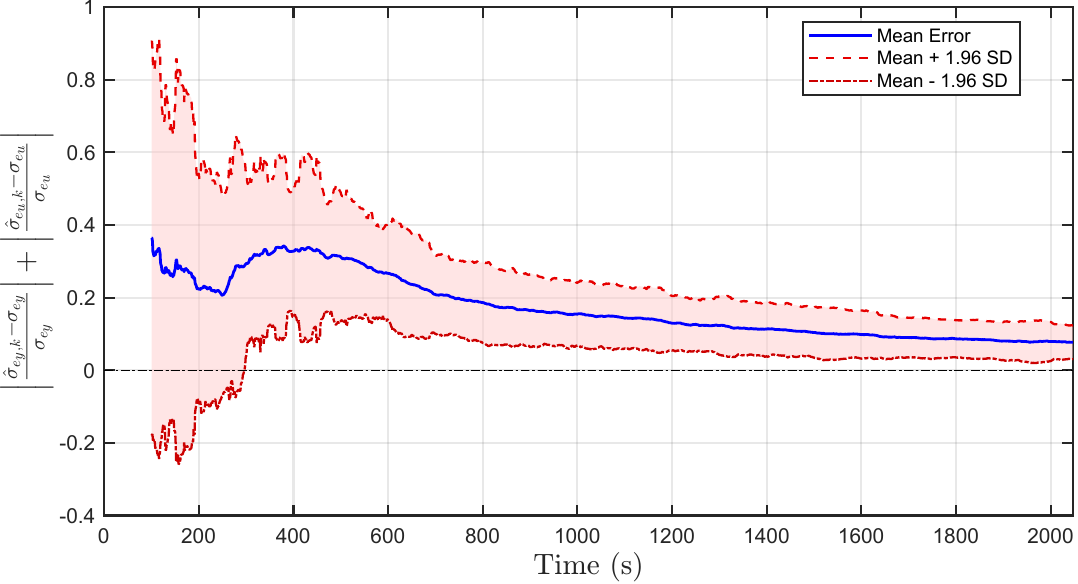}    
    \caption{Sum of the relative errors in the estimates of noise standard deviations, computed at $k = 100, \ldots, 2047$.} 
    \label{Figure_5}
    \end{center}
\end{figure}

We perform $50$ simulation tests and report the mean and $95\%$ confidence intervals of the two performance metrics specified earlier, in  Figs.~\ref{Figure_4} and \ref{Figure_5}, respectively.

It can be observed from Fig.~\ref{Figure_4} that the process order converges to true process order $\eta = 2$ within $160$ samples. Subsequently the process order is correctly estimated even though the noise variances gradually increase from the $250$'th time instant. We further observe from Fig.~\ref{Figure_5} that the error in the estimated noise variances gradually decrease until the $250$'th instant, after which it initially increases due to induced changes in the sensor noise variances.  However, as more measurements are received, the estimates of the noise variances gradually converge to their true values. The final estimates of the noise variances obtained are $\hat{\sigma}_{e_y}^2 = 0.439 \pm 0.062$ and $\hat{\sigma}_{e_u}^2 = 0.189 \pm 0.046$, which are close to the simulated values. The estimated model coefficients, averaged over the $50$ simulation trials along with their respective $95\%$ CI at the end of the simulation run ($2047$'th time instant) are given below
\begin{align}
    \label{eq:4.2}
        &y^*(k) = \underset{\pm (0.046)}{1.489}y^*(k-1) - \underset{\pm (0.047)}{0.689}y^*(k-2)\ + \\ & \underset{\pm (0.034)}{0.002}u^*(k) + \underset{\pm (0.050)}{1.002}u^*(k-1) + \underset{\pm (0.075)}{0.514}u^*(k-2) \nonumber
\end{align}
The results indicate that rARX-DIPCA is able to obtain unbiased estimates of the noise variances, and the model parameters even at a low SNR. Although the modified DIPCA method is developed under the assumption that the noise variances is constant over all samples, rARX-DIPCA is able to track sensor degradation as new samples are received, due to re-estimating the noise variances at every time instant.

\subsection{Tracking Changes in Operating Conditions} \label{subsec:study2}
For this case study, we consider a second-order system with third-order input dynamics as described in the following
\begin{equation}
    \begin{array}{ll}
        \label{eq:4.3}
        y^*(k) =& 1.1y^*(k-1) - 0.7y^*(k-2)\ + \\& u^*(k-2) + 0.5u^*(k-3)
    \end{array}
\end{equation}
A nonlinear process can be modeled using an approximate linear model with time varying model coefficients to capture the process behavior induced by changes in operating conditions. In order to assess the capability of the proposed rARX-DIPCA method for such an application, in this simulation study the model coefficients are varied. We first choose full band RBS as input $u^*$ with sample size $6095$, and true values of $y^*$ are generated using (\ref{eq:4.3}) for first $400$ samples, after which we vary the coefficients in quadratic fashion till $650$'th instant, yielding the following updated system for the remaining instances
\begin{equation}
    \begin{array}{ll}
        \label{eq:4.4}
        y^*(k) =& y^*(k-1) - 0.6y^*(k-2)\ + \\& 1.2u^*(k-2) + 0.7u^*(k-3)
    \end{array}
\end{equation}
Noise is added to the true values following the ARX model to generate the measurements, using noise variances $\sigma^2_{e_u} = 0.1$ and $\sigma^2_{e_y} = 0.15$, which correspond to SNR $10$. Fifty simulation trials are conducted with different noise realizations, each with $6095$ measurement observations. 

We obtain the lagged sample covariance, along with the estimates of the input-output noise variances, process order, and model coefficients by applying the modified Dynamic IPCA considering the first $100$ samples, to initialize the rARX-EIV method. At each subsequent sampling instant, upon receiving new measurements, the proposed recursive identification algorithm is applied to re-estimate all the model parameters. The performance metrics described in preceding sub-section are again used to assess the performance of the method.

\begin{figure}[!htbp]
    \begin{center}
    \includegraphics[width=8.4cm]{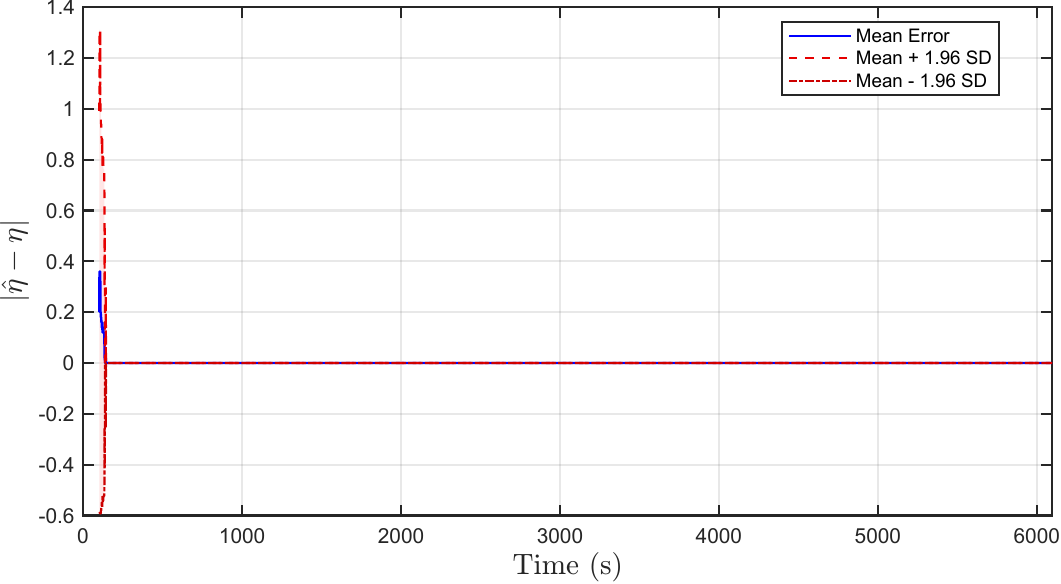}    
    \caption{Absolute difference between the estimated and true process orders.} 
    \label{Figure_6}
    \end{center}
\end{figure}

\begin{figure}[!htbp]
    \begin{center}
    \includegraphics[width=8.4cm]{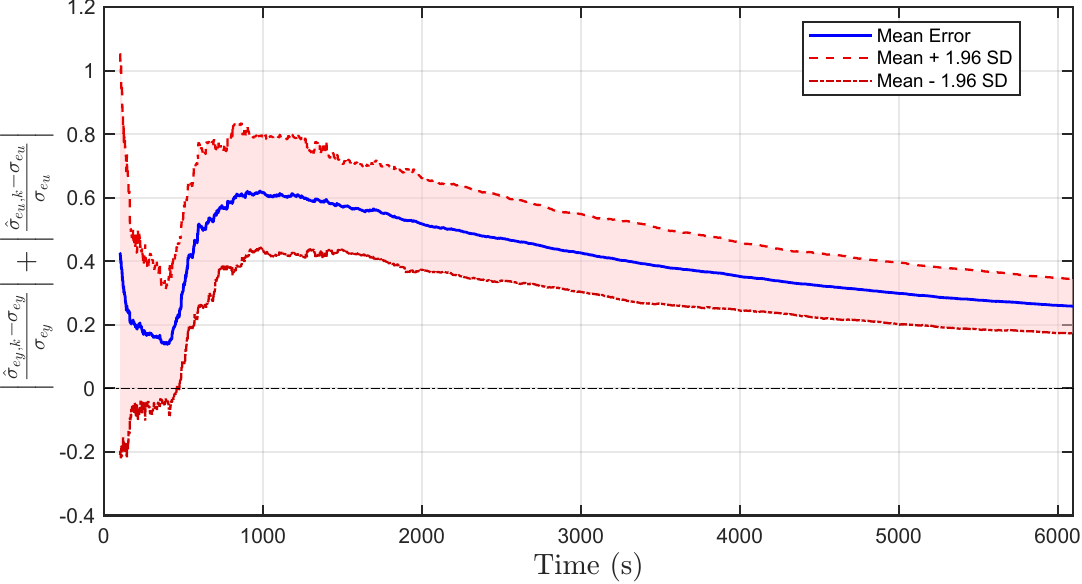}    
    \caption{Sum of the relative errors in the estimates of noise standard deviations, computed at $k = 100, \ldots, 6095$.} 
    \label{Figure_7}
    \end{center}
\end{figure}

From Fig.~\ref{Figure_6}, we observe that the estimate of the process order converges to the true value $\eta = 3$ within $200$ samples. Fig.~\ref{Figure_7} shows that the error in the estimated noise variances decreases before the changes in the model coefficients are introduced. Although the noise variances are kept constant, the error in the estimated noise variances increases for a short period after the $650$'th time instant when changes in the model coefficients are introduced.  This is due to the fact that the linear model identification is developed under the assumption that the entire sample of measurements used for model identification is generated from a constant time invariant linear process.  Despite this, the estimated noise variances start converging to their respective true values as additional measurements corresponding to the modified process are received. The noise variance estimates obtained at the end of the simulation are $\hat{\sigma}^2_{e_u} = 0.089 \pm 0.014$, and $\hat{\sigma}^2_{e_y} = 0.173 \pm 0.027$, which indicate that they are unbiased. Since accurate estimation of the noise variances is the primary requirement for unbiased PCA-based parameter estimation, the successful tracking of these variances, as illustrated in Fig.~\ref{Figure_7}, also serves as strong evidence of the algorithm's parameter-tracking capabilities throughout the transition period.

\begin{table}[!htbp]
\centering
\caption{Comparing the model coefficient estimates of the systems defined in (\ref{eq:4.3}) and (\ref{eq:4.4}).}
\label{table:1}
\begin{minipage}{8.34cm}
\centering
\begin{threeparttable}
    \renewcommand{\arraystretch}{1.35}
    \begin{tabular}{ccccc}
    \hline
    & \multicolumn{2}{c}{\makecell{Estimation before \\ $401$'th instance}} & \multicolumn{2}{c}{\makecell{Estimation over \\ $6095$ observations}} \\
    \cmidrule(lr){2-3} \cmidrule(lr){4-5}
    \makecell{Corresponding \\ Variable} & \makecell{True \\ values} & \makecell{Estimated \\ values} & \makecell{True \\ values} & \makecell{Estimated \\ values}  \\
    \hline
    $y^*(k)$ & $1.0$ & $1.0$ & $1.0$ & $1.0$ \\
    $y^*(k-1)$ & $1.1$ & $\underset{\pm (0.038)}{1.117}$ & $1.0$ & $\underset{\pm (0.024)}{1.023}$ \\
    $y^*(k-2)$ & $-0.7$ & $\underset{\pm (0.050)}{-0.724}$ & $-0.6$ & $\underset{\pm (0.028)}{-0.628}$ \\
    $y^*(k-3)$ & $0.0$ & $\underset{\pm (0.032)}{0.017}$ & $0.0$ & $\underset{\pm (0.019)}{0.013}$ \\
    $u^*(k)$ & $0.0$ & $\underset{\pm (0.025)}{0.008}$ & $0.0$ & $\underset{\pm (0.016)}{0.002}$ \\
    $u^*(k-1)$ & $0.0$ & $\underset{\pm (0.029)}{0.011}$ & $0.0$ & $\underset{\pm (0.012)}{-0.0004}$ \\
    $u^*(k-2)$ & $1.0$ & $\underset{\pm (0.025)}{1.007}$ & $1.2$ & $\underset{\pm (0.011)}{1.173}$ \\
    $u^*(k-3)$ & $0.5$ & $\underset{\pm (0.051)}{0.487}$ & $0.7$ & $\underset{\pm (0.033)}{0.659}$ \\
    \hline
    \end{tabular}
\end{threeparttable}
\end{minipage}
\end{table}

The model coefficient estimates, obtained before the $401$'th instant (before the changes in the model coefficients are introduced), and at the final sampling instant $(\text{i.e.,}\ k=6095)$ are reported in table~\ref{table:1} along with their respective $95\%$ confidence intervals. The results indicate that rARX-DIPCA has obtained unbiased estimates of the initial process model coefficients as defined in (\ref{eq:4.3}), and has also adapted to the time-varying process by providing unbiased estimates of the final process model coefficients given by (\ref{eq:4.4}). However, the estimated coefficients at the $6095$'th instant, corresponding to the variables $u^*(k-2)$ and $u^*(k-3)$ have very small bias, which, as per the observed convergence trend, is expected to vanish as more measurements are received.

\section{Conclusion}\label{sec:conclusion}
This paper has introduced rARX-DIPCA, a recursive identification algorithm for time-varying EIV-ARX processes without the need for complete historical data storage. Simulation studies show that the proposed method is able to adapt to changes in the noise variances due to sensor degradation, and model coefficient changes resulting from changing operating conditions. Future research directions potentially include extending the framework to handle multi-variable MIMO-ARX systems in order to address more complex industrial processes with coupled variables, incorporating time-varying delay tracking capabilities to handle systems with transport phenomena or communication latency. Theoretical convergence analysis under persistent parameter variations would strengthen the method's foundation for safety-critical applications.
 

\begin{thebibliography}{15}
\providecommand{\natexlab}[1]{#1}
\providecommand{\url}[1]{\texttt{#1}}
\providecommand{\urlprefix}{URL }
\expandafter\ifx\csname urlstyle\endcsname\relax
  \providecommand{\doi}[1]{doi:\discretionary{}{}{}#1}\else
  \providecommand{\doi}{doi:\discretionary{}{}{}\begingroup \urlstyle{rm}\Url}\fi

\bibitem[{Cao et~al.(2018)Cao, Luo, and Song}]{Cao:2018}
Cao, P., Luo, X., and Song, X. (2018).
\newblock Modeling and identification for soft sensor systems based on the separation of multi-dynamic and static characteristics.
\newblock \emph{Chinese Journal of Chemical Engineering}, 26(1), 137--143.

\bibitem[{Diversi et~al.(2010)Diversi, Guidorzi, and Soverini}]{Diversi:2010}
Diversi, R., Guidorzi, R., and Soverini, U. (2010).
\newblock Identification of {ARX} and {ARARX} models in the presence of input and output noises.
\newblock \emph{European Journal of Control}, 16(3), 242--255.

\bibitem[{Fernando and Nicholson(1985)}]{Fernando:1985}
Fernando, K. and Nicholson, H. (1985).
\newblock Identification of linear systems with input and output noise: the {Koopmans-Levin} method.
\newblock \emph{IEE Proceedings D (Control Theory and Applications)}, 132, 30--36.

\bibitem[{Ikenoue et~al.(2005)Ikenoue, Kanae, Yang, and Wada}]{Ikenoue:2005}
Ikenoue, M., Kanae, S., Yang, Z.J., and Wada, K. (2005).
\newblock Identification of noisy input-output system using bias-compensated least-squares method.
\newblock \emph{IFAC Proceedings Volumes}, 38(1), 803--808.
\newblock 16th IFAC World Congress.

\bibitem[{Li et~al.(2000)Li, Yue, Valle-Cervantes, and Qin}]{Li:2000}
Li, W., Yue, H., Valle-Cervantes, S., and Qin, S. (2000).
\newblock Recursive {PCA} for adaptive process monitoring.
\newblock \emph{Journal of Process Control}, 10(5), 471--486.

\bibitem[{Maurya et~al.(2018)Maurya, Tangirala, and Narasimhan}]{Maurya:2018}
Maurya, D., Tangirala, A.K., and Narasimhan, S. (2018).
\newblock Identification of errors-in-variables models using dynamic iterative principal component analysis.
\newblock \emph{Industrial \& Engineering Chemistry Research}, 57(35), 11939--11954.

\bibitem[{Maurya et~al.(2022)Maurya, Tangirala, and Narasimhan}]{Maurya:2022}
Maurya, D., Tangirala, A.K., and Narasimhan, S. (2022).
\newblock Identification of errors-in-variables {ARX} models using modified dynamic iterative {PCA}.
\newblock \emph{Journal of the Franklin Institute}, 359(13), 7069--7090.

\bibitem[{Picci et~al.(2023)Picci, Falconi, Ferrante, and Zorzi}]{Picci:2023}
Picci, G., Falconi, L., Ferrante, A., and Zorzi, M. (2023).
\newblock Hidden factor estimation in dynamic generalized factor analysis models.
\newblock \emph{Automatica}, 149, 110834.

\bibitem[{Pradeep and Narasimhan(2025)}]{Pradeep:2025}
Pradeep, M. and Narasimhan, S. (2025).
\newblock Recursive iterative principal component analysis.
\newblock \emph{Computers \& Chemical Engineering}, 201, 109170.

\bibitem[{Prakash et~al.(2005)Prakash, Narasimhan, and Patwardhan}]{Prakash:2005}
Prakash, J., Narasimhan, S., and Patwardhan, S.C. (2005).
\newblock Integrating model based fault diagnosis with model predictive control.
\newblock \emph{Industrial \& Engineering Chemistry Research}, 44(12), 4344--4360.

\bibitem[{Prakash and Huang(2025)}]{Prakash:2025}
Prakash, O. and Huang, B. (2025).
\newblock Real-time update of data-driven reduced and full order models with applications.
\newblock \emph{Computers \& Chemical Engineering}, 194, 108923.

\bibitem[{S{\"o}derstr{\"o}m and Stoica(2002)}]{Soderstrom:2002}
S{\"o}derstr{\"o}m, T. and Stoica, P. (2002).
\newblock Instrumental variable methods for system identification.
\newblock \emph{Circuits, Systems and Signal Processing}, 21(1), 1--9.

\bibitem[{Zhang et~al.(2022)Zhang, Guo, Hao, Chen, and Huang}]{Zhang:2022}
Zhang, J., Guo, F., Hao, K., Chen, L., and Huang, B. (2022).
\newblock Identification of errors-in-variables {ARX} model with time varying time delay.
\newblock \emph{Journal of Process Control}, 115, 134--144.

\bibitem[{Zhao et~al.(2018)Zhao, Fatehi, and Huang}]{Zhao:2018}
Zhao, Y., Fatehi, A., and Huang, B. (2018).
\newblock Robust estimation of {ARX} models with time varying time delays using {V}ariational {B}ayesian approach.
\newblock \emph{IEEE Transactions on Cybernetics}, 48(2), 532--542.

\bibitem[{Zheng(2002)}]{Zheng:2002}
Zheng, W.X. (2002).
\newblock A bias correction method for identification of linear dynamic errors-in-variables models.
\newblock \emph{IEEE Transactions on Automatic Control}, 47(7), 1142--1147.

\end{thebibliography}


\end{document}